\pdfoutput=1
\documentclass[aps,prl,floatfix,twocolumn,showpacs,preprintnumbers,amsmath,amssymb,a4paper,superscriptaddress]{revtex4-1}
\usepackage{braket}
\usepackage{dsfont}
\usepackage{amsmath}
\usepackage{amssymb}
\usepackage{mathtools}
\usepackage{graphicx}% Include figure files
\usepackage{dcolumn}% Align table columns on decimal point
\usepackage{bm}% bold math
\usepackage{multirow}
\usepackage{leftidx}
\usepackage{color}
\usepackage[german,english]{babel}

\begin{document}

\title{Ultracold heteronuclear three-body systems: How diabaticity limits the universality of recombination into shallow dimers}

 \author{P. Giannakeas}
 \email{pgiannak@purdue.edu}
 \affiliation{Department of Physics and Astronomy, Purdue University, West Lafayette, Indiana 47907, USA}
 \affiliation{Max Planck Institute for the Physics of Complex Systems, N\"othnitzer Strasse 38, 01187 Dresden, Germany}
 
 \author{Chris H. Greene}
 \email{chgreene@purdue.edu}
 \affiliation{Department of Physics and Astronomy, Purdue University, West Lafayette, Indiana 47907, USA}
 
 \date{\today}

\begin{abstract}
The mass-imbalanced three-body recombination process that forms a shallow dimer is shown to possess a rich Efimov-St\"uckelberg landscape, with corresponding spectra that differ fundamentally from the homonuclear case.
A semi-analytical treatment of the three-body recombination predicts an unusual spectra with intertwined resonance peaks and minima, and yields in-depth insight into the behavior of the corresponding Efimov spectra.
In particular, the patterns of the Efimov-St\"uckelberg landscape are shown to depend inherently on the degree of diabaticity of the three-body collisions, which strongly affects the universality of the heteronuclear Efimov states.
\end{abstract}


\maketitle
The infinite geometric progression of trimer states, even when all the two-body subsystems are barely unbound
consists of the most counterintuitive few-body phenomenon, namely the {\it Efimov effect} \cite{v_efimov_hard-core_1970, kraemer_evidence_2006}.
The exotic Efimov states have received extensive theoretical and experimental study \cite{nielsen_three-body_2001,riisager_nuclear_1994,braaten_universality_2006,rittenhouse_hyperspherical_2011,blume_few-body_2012,wang_chapter_2013,wang2015few,naidon_review_2017,greene_review_2017} addressing the underlying physical principles, such as the universality of the ground Efimov state \cite{wang_universal_2014-1,roy_prl_2013,wang_origin_2012,gross_observation_2009,naidon_microscopic_2014,naidon_physical_2014,ferlaino_forty_2010,kunitski_observation_2015}, and the discrete scaling invariance for successive Efimov states for homonuclear \cite{huang_observation_2014} and heteronuclear systems, i.e. heavy-heavy-light (HHL), \cite{pires_observation_2014prl,ulmanis_universality_2015,tung_geometric_2014-1}.
Recent experimental evidence on mass-imbalanced ensembles, suggests that the Efimov spectra possess a far richer landscape than the homonuclear counterparts \cite{ulmanis_heteronuclear_2016,hafner2017role,johansen_testing_2017} stemming from the large parameter space.

HHL systems possess two scattering lengths, i.e. $a_{HH}$ and $a_{HL}$ which define four main categories of behavior according to the signs alone since each $a_{ij}$ can be positive or negative.
Additional sub-categories unfold depending on magnitude, i.e. $|a_{HH}|/|a_{HL}|\gtrless 1$.
Spanning a large portion of parameter space, Refs.\cite{pires_observation_2014prl, ulmanis_universality_2015,hafner2017role,petrov_three-body_2015,mikkelsen_three-body_2015} experimentally and theoretically explored the three-body losses of $^6$Li-$^{133}$Cs-$^{133}$Cs system demonstrating that the different signs in the intraspecies scattering lengths ($a_{\rm{CsCs}}\gtrless 0$) render an inherently different Efimovian landscape.
Additional experimental efforts for the case $a_{HH}>0$ illustrate deviations of the universal theory from the observed Efimov spectra \cite{johansen_testing_2017} in the regime where the two-body interspecies interactions, i.e. $a_{HL}$ are tuned via narrow Fano-Feshbach resonances \cite{chin_feshbach_2010}.
These investigations pose the most intriguing questions in the few-body physics of HHL systems: whether the three-body physics is universal, whether the Efimov spectrum needs a nonuniversal ``3-body parameter'' (3BP) to specify its lowest state, and whether van der Waals (vdW) physics approximately determines that 3BP, as experimental and theoretical evidence suggests is true for the homonuclear case \cite{ferlaino_forty_2010,gross_observation_2009-1,roy_prl_2013,wang_origin_2012,naidon_microscopic_2014,naidon_physical_2014}.
Therefore, a more flexible and complete theoretical description of three-body recombination (3BR) into shallow dimers ($a_{HH}>0$) is needed.

This Letter develops a semiclassical theoretical treatment based on the adiabatic hyperspherical representation with nonadiabatic coupling included addressing shallow dimer recombination for HHL systems. 
We demonstrate that the universality for various observables in this system is strongly affected by the {\it degree of diabaticity} connecting the three-body continuum and recombination channels.
Adiabatic collisions exhibit the Efimov physics idiosyncrasies which depend {\it only} on the scattering lengths, rendering the three-body system {\it fully} universal.
In contrast from the homonuclear case, heteronuclear systems with positive intraspecies and negative interspecies interactions (i.e. $a_{\rm{HH}}>0$ and $a_{\rm{HL}}<0$, respectively) exhibit 3BR rates whose landscape for varying scattering lengths was previously predicted \cite{dincao_ultracold_2009} to consist of Efimov resonances intertwined with a series of {\it{St\"uckelberg interference minima}} \cite{nielsen_low-energy_1999,esry1999recombination}.
The present analysis shows how the assumption of hyperradial adiabaticity underlying the original prediction by D'Incao and Esry\cite{dincao_ultracold_2009} must be generalized when the relevant Landau-Zener transition probability is closer to diabatic than adiabatic, as is true for strong mass-imbalanced systems.

Our prototype three-body system consists of two heavy (H) atoms and a light (L) one, which collide at low energies via $s$-wave {\it{zero-range}\/} interactions.
 The 3BR into a shallow heavy-heavy (HH) dimer and a recoiling light atom is achieved by considering the intraspecies (interspecies) interactions to possess positive (negative) scattering length, i.e. $a_{\rm{HH}}>0$ ($a_{\rm{HL}}<0$).
 The paraphernalia of the {\it{adiabatic hyperspherical representation} \/} (for a detailed review see \cite{greene_review_2017}) is employed in the following addressing the three-body physics of interest.
In this representation the properly symmetrized total wavefunction is written as $\Psi(R,\Omega)=R^{-5/2}\sum_\nu \phi_\nu(R;\Omega)F_\nu(R)$, where $\phi_\nu(R;\Omega)$ [$F_\nu(R)$] refers to the $\nu$-th hyper-angular [hyper-radial] component of $\Psi(R,\Omega)$.  The hyperangular factor is an eigenfunction of the fixed-$R$ adiabatic equation \cite{suno_three-body_2003}:
\begin{equation}
  H_{\rm{ad}}(R;\Omega) \phi_\nu(R;\Omega)=U_\nu(R)\phi_\nu(R;\Omega),
  \label{eq:eq0}
\end{equation}
where $\Omega$ collectively denotes the five {\it{hyperangles}\/} whereas $R$ is the {\it{hyperradius}\/} \cite{avery_hyperspherical_1989,smirnov_method_1977}.
The $U_\nu(R)$ are the {\it{adiabatic hyperspherical potential curves}\/} and $H_{\rm{ad}}$ contains the hyperangular kinetic operator together with the pairwise zero-range interactions.
After integrating over all the hyperangles $\Omega$, the three-body Schr\"odinger equation reads:

\begin{equation}
  \bigg[-\frac{d^2}{d R^2}+\frac{2 \mu}{\hbar^2}(U_\nu-E)\bigg]F_\nu(R)= \sum_{\nu '}V_{\nu \nu'}F_{\nu '}(R),
  \label{eq:eq1}
\end{equation}

where $\mu=m_H/\sqrt{1+2m_H/m_L}$ is the three-body reduced mass, $m_H$ ($m_L$) denotes the mass of the heavy (light) atom  and $E$ indicates the total relative energy.
$V_{\nu \nu'}=2P_{\nu \nu '}(R)\frac{d}{dR}+ Q_{\nu \nu'}(R)$ are the $R$-dependent {\it{non-adiabatic} \/} coupling matrix elements obeying the relations: $P_{\nu \nu'}=\braket{\phi_\nu(R; \Omega)|\frac{\partial}{\partial R}\phi_{\nu '}(R;\Omega)}_{\Omega}$ and $Q_{\nu \nu'}=\braket{\phi_\nu(R; \Omega)|\frac{\partial^2}{\partial R^2}\phi_{\nu '}(R;\Omega)}_{\Omega}$.
Note that the symbol $\braket{\ldots}_\Omega$ indicates the integration over the hyper-angles {\it{only} \/}.
For zero-range interactions $U_\nu$, $P_{\nu \nu'}$ and $Q_{\nu \nu'}$ are known semi-analytically \cite{rittenhouse_greens_2010,kartavtsev_low-energy_2007,kartavtsev_universal_2006,nielsen_three-body_2001}.

%    \pagebreak
\begin{figure}[t]
\centering
 \includegraphics[scale=0.33]{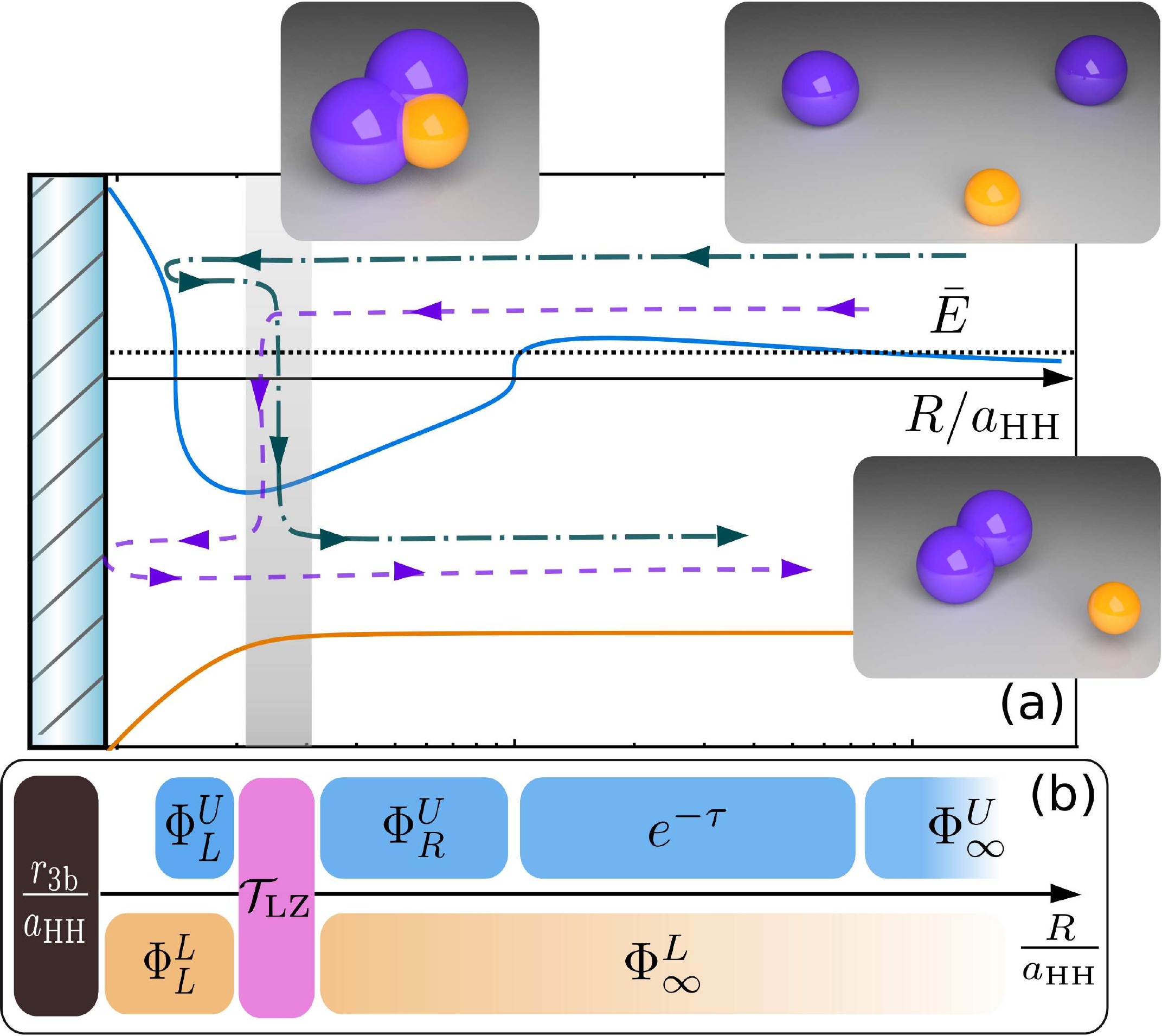}
 \caption{(Color on line) (a) The two lowest hyperspherical potential curves $U^{1/3}_\nu(\frac{R}{a_{HH}})$ (orange and blue solid curves) for a HHL system with mass ratio $m_H/m_L=22.1$, $a_{HL}<0$, $a_{HH}>0$ and the total energy $\bar{E}$ in units of $\frac{\hbar^2}{m_{H} a_{HH}^2}$ is indicated as a black dotted line.
   The blue hatched area indicates the 3BP at $R=r_{3b}$ whereas the gray region depicts the maximum of the $P$-matrix. The green dash-dotted and purple dashed lines denote the dominant two interfering pathways.
 (b) An illustration of the terms associated with the semiclassical treatment and their connections to the curves shown in panel (a). }
\label{fig1}
\end{figure}

Fig.\ref{fig1}(a) illustrates within the two-channel approximation the physical picture of a shallow dimer recombination process with $a_{HH}>0$ and $a_{HL}<0$.
In Fig.\ref{fig1}(a) the first two adiabatic potential curves $U_\nu^{1/3}(R)$ (solid blue and orange lines) are depicted as functions of the scaled hyper-radius $R/a_{HH}$. 
The mass ratio between the heavy and light particle is set $m_{H}/m_{L}=22.1$ corresponding to $~^{6} {\rm Li}-^{133} {\rm Cs}-^{133}{ \rm{Cs}}$ collisions.
The 3BP is introduced as a hard wall boundary condition at $R=r_{\rm{3b}}$ [see blue hatched area in Fig.\ref{fig1}(a)], avoiding the {\it{Thomas collapse}} \cite{thomas_interaction_1935-1}.
The gray box in Fig.\ref{fig1}(a) indicates the hyper-radial region where the non-adiabatic coupling $P$-matrix element maximizes, namely the {\it{non-adiabatic transition region}}.
At the indicated total relative energy $\bar{E}$, written in units of $\frac{\hbar^2}{m_H a_{\rm{HH}}^2}$ [see dotted line in Fig.\ref{fig1}(a)] and $R/a_{\rm{HH}}\to\infty$ the three particles are asymptotically free.  
But as $R/a_{\rm{HH}}$ decreases, the system tunnels inward under the barrier probing the classically-allowed region of the upper potential curve (blue solid line).
Due to the non-adiabatic coupling (see gray box) the three particles can recombine to the lower potential curve (orange solid line) forming a shallow HH dimer and a recoiling light particle.
To quantitatively address this physical process, a fully semiclassical treatment of Eq.~(\ref{eq:eq1}) is developed within the two-channel approximation, yielding an analytical expression for the corresponding 3BR rate $K_3$:

\begin{eqnarray}
  K_3&=&\frac{64 \hbar \pi^2}{\mu k^4}|S_{12}|^2,~{\rm{with}}~|S_{12}|^2=e^{-2\tau} p(1-p)\frac{N}{D},\nonumber \\
  N&=&\cos^2(\Phi^U_L-\Phi^L_L-\frac{\pi}{4}+\lambda),~~D=(1-\frac{e^{-4\tau}}{16})[p\times \nonumber \\
  &\times&\cos^2(\Phi_L^L+\Phi^U_R-\frac{\pi}{4})+(1-p)\cos^2(\Phi_L^U+\Phi^U_R+\lambda)]\nonumber \\
  &-&(1-\frac{e^{-2\tau}}{4})^2 p(1-p)N+\frac{e^{-4\tau}}{16}.
  \label{eq:eq2}
\end{eqnarray}

Here $k^2=2\mu E/\hbar^2$ and $|S_{12}|^2$ denotes the $S$-matrix element describing the 3BR into a universal shallow dimer.
Fig.\ref{fig1}(b) depicts the terms that appear in Eq.~(\ref{eq:eq2}).
The terms $\Phi^i_\alpha$ with $i=L, U$ and $\alpha=L,R$ indicate the JWKB phases (including the Langer correction \cite{nielsen_low-energy_1999}) of the upper ($i=U$) and lower ($i=L$) potential curves of Fig.\ref{fig1}(a) on the left ($\alpha=L$) and right hand side ($\alpha=R$) of the transition region, i.e.\  the gray box of Fig.\ref{fig1}(a).
The asymptotic phases of the upper and lower curves are indicated by $\Phi_\infty^U$ and $\Phi_\infty^L$, respectively, and of course the $|S_{12}|^2$ does not depend on them.
Note that  $\Phi_\infty^U$ is defined from the outer classical turning point of the upper Efimov curve whereas $\Phi_\infty^L$ is defined from hyperradii beyond the non-adiabatic transition region.
In Fig.\ref{fig1}(b) the factor $e^{-\tau}$ indicates the tunneling amplitude in the classically-forbidden region of the upper potential curve [see the blue solid line in Fig.\ref{fig1}(a)].
The 3BP is indicated in Fig.\ref{fig1}(b) by the far left black box.

The phase $\lambda$, and the term $p$ in Eq.~(\ref{eq:eq2}) as well as the term $\mathcal{T}_{LZ}$ in Fig.\ref{fig1}(b) are associated with the Landau-Zener physics \cite{note1}.
$p$ indicates the non-adiabatic transition probability from the upper to the lower potential curve shown in Fig.\ref{fig1}(a) which is evaluated by the $P$-matrix elements \cite{clark_calculation_1979}.
The phase $\lambda$ is associated with the pass of the hyper-radial wavefunction through the non-adiabatic transition region and solely depends on the probability $p$ \cite{child1974semiclassical,zhu1994theory}. 
This non-trivial phase is necessary for an accurate 3BR coefficient, as our numerical tests suggest.
The $\mathcal{T}_{LZ}\equiv\mathcal{T}_{LZ}(\lambda, p)$ denotes the non-adiabatic transition matrix which inter-relates the adiabatic hyper-radial wavefunction bilaterally of the transition region as shown in Fig.\ref{fig1}(a) \cite{child1974semiclassical,zhu1994theory}. 

\begin{figure}[t]
  \centering
  \includegraphics[scale=0.5]{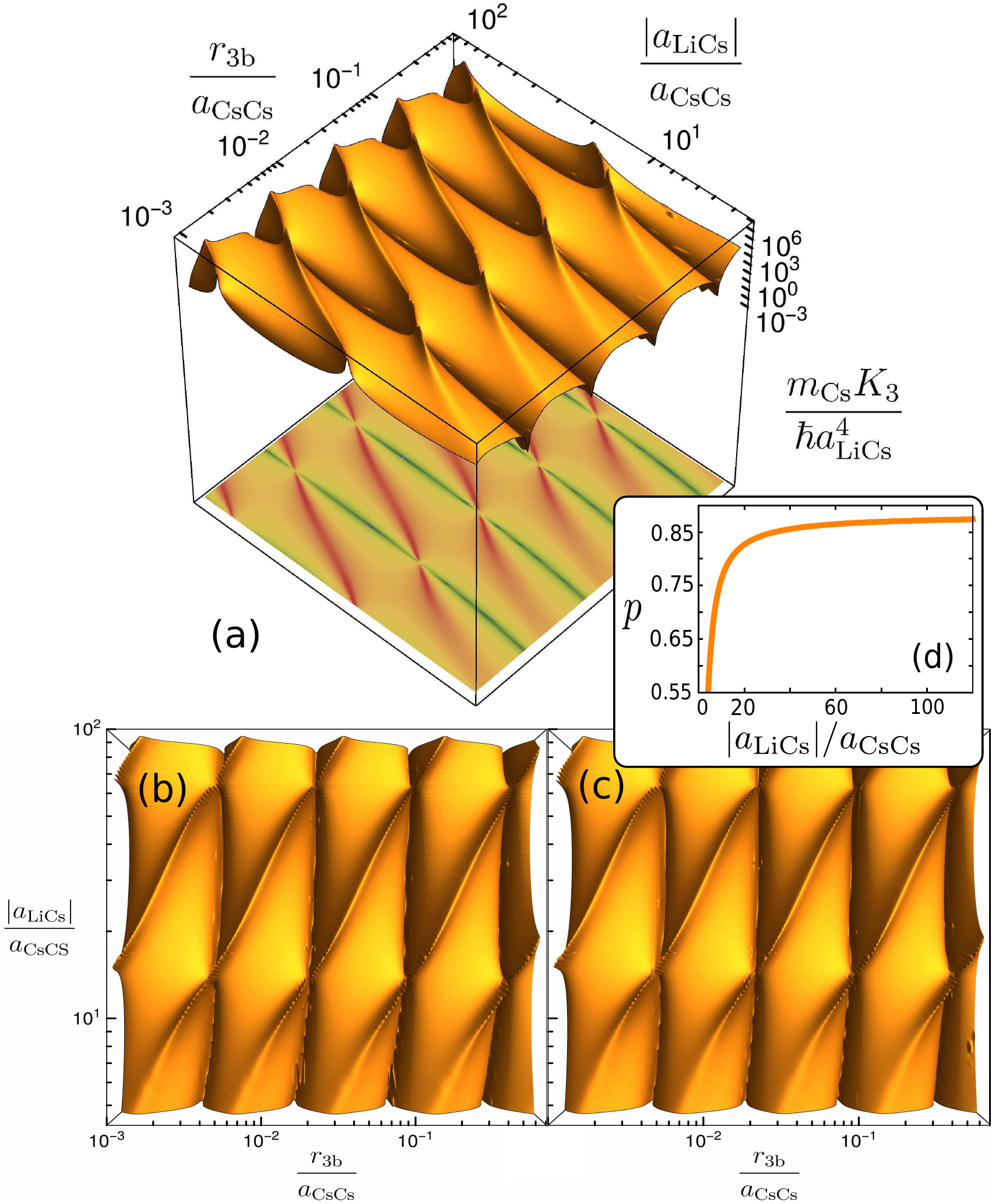}
  \caption{(color online) The scaled 3BR coefficient $\frac{m_{\rm{Cs}}K_3}{\hbar a_{\rm{LiCs}}^4}$ at low energy $\bar{E}$ (in units of $\frac{\hbar^2}{m_{H} a_{HH}^2}$) for $~^6 {\rm{Li}}-^{133}{\rm{Cs}}-^{133}{\rm{Cs}}$ system as a function of two ratios: the 3BP $r_{\rm{3b}}/a_{\rm{CsCs}}$ and the interspecies scattering length $|a_{\rm{LiCs}}|/a_{\rm{CsCs}}$. Panels (b)-(c) show the scaled 3BR calculations within the semiclassical treatment and the R-matrix numerical simulations, respectively. Panel (d) depicts the probability $p$ versus $|a_{\rm{LiCs}}|/a_{\rm{CsCs}}$. }
\label{fig2}
\end{figure}

The analytical expression of $K_3$ in Eq.~(\ref{eq:eq2}) conveys the most important attributes of a recombination process into a shallow HH dimer for the HHL system.
The key role is played here by the {\it degree of {diabaticity}\/}, namely the non-adiabatic probability $p$ in $K_3$ and inherently modifies the properties of the Efimov spectra whereas in the homonuclear cases the corresponding $p$ is trivially an overall pre-factor in $K_3$. 
The resonant enhancement of $K_3$ mainly arises due to the resonant denominator $D$ in Eq.~(\ref{eq:eq2}) which explicitly depends on the non-adiabatic transition probability $p$.
Hence, the Efimov spectra are influenced by the degree of diabaticity of the three-body collisions.
For example, in a mostly diabatic collisions, i.e.\ $p\approx 1$, the Efimov resonances in $K_3$ depend on the phase $\Phi^L_L$ which is defined by the 3BP [see Fig.\ref{fig1}(b)].
For adiabatic three-body collisions, i.e.\ $p\approx 0$, $K_3$ is virtually independent of the $\Phi_L^L$ phase.
Therefore, a mostly adiabatic 3BR process possesses a {\it fully universal} Efimov spectrum which is independent of any 3BP.
This adiabatic limit was stressed in the case of vdW two-body interactions study presented by Refs. \cite{ulmanis_heteronuclear_2016, hafner2017role}.
The probability $p$ in general depends mainly on the $\frac{|a_{HL}|}{a_{HH}}$ and the mass ratio $\beta=\frac{m_H}{m_L}$ but in principle it could be affected by the vdW physics in realistic scenarios.
 In Fig.\ref{fig2}(d) $p$ varies monotonically in the interval $(0.55,0.87)$ for $\frac{|a_{HL}|}{a_{HH}}\in(4.2,123.4)$, respectively, matching the regime of Refs.\cite{ulmanis_heteronuclear_2016, hafner2017role,johansen_testing_2017} around $B=889~\rm{G}$.
In the unitarity limit ($\frac{|a_{HL}|}{a_{HH}}\to \infty$), $p$ obeys the fitting relation $p_{\infty}\sim e^{-\frac{16}{2\beta+1}+0.22}$ for $2.3<\beta<23$ to a good approximation.
The latter implies that large mass ratios yield mostly diabatic 3BR processes.

Eq.~(\ref{eq:eq2}) exhibits also signatures of St\"uckelberg physics.
The numerator of $|S_{12}|^2$ depends on the phases $\Phi_L^U$ and $\Phi_L^L$ emerging from a two-pathway interference [see green dash-dotted and purple dashed lines in Fig.\ref{fig1}(a)] and can cause a suppression of $K_3$, i.e. {\it{St\"uckelberg minima}} which is non-universal due to the $\Phi_L^L$ phase.
Note that the semiclassical theory developed here for $K_3$ in Eq.~(\ref{eq:eq2}) generalizes and extends the study by Ref.\cite{dincao_ultracold_2009}, through our inclusion of the explicit dependence on the non-adiabatic probability $p$.
This analysis also  provides a systematic pathway to generalize Eq.~(\ref{eq:eq2}) to positive inter- and intraspecies interactions going beyond previous studies \cite{helfrich_three_body_2010, acharya_pra_2016}.

Our calculated 3BR coefficient is illustrated in Fig.\ref{fig2} where the system of $~^6 {\rm{Li}}-^{133}{\rm{Cs}}-^{133}{\rm{Cs}}$ is considered.
Fig.\ref{fig2} depicts the scaled 3BR coefficient,  $\frac{m_{\rm{Cs}} K_3}{\hbar a_{\rm{LiCs}}^4}$, at low energies $\bar{E}$ (in units of $\frac{\hbar^2}{m_{H} a_{HH}^2}$), as a function of the ratios $\frac{r_{\rm{3b}}}{a_{\rm{CsCs}}}$ and $\frac{|a_{\rm{LiCs}}|}{a_{\rm{CsCs}}}$.
Fig.\ref{fig2}(a) refers to a direct numerical solution of Eq.~(\ref{eq:eq1}) within the two-channel approximation using the R-matrix propagation method (for details see Refs.\cite{mehta_three-body_2007,burke_jr_theoretical_1999}).
Panel(a) conveys the rich structure that arises in the 3BR coefficient.
The enhancements in $\frac{K_3}{a^4_{\rm{LiCs}}}$ emerge from metastable Efimov states that resonantly transfer probability flux from the three-body continuum into the atom-dimer continuum via the non-adiabatic region (see the red stripes in the contour plot of Fig.\ref{fig2}(a)).
Within the zero-range approximation the position of the Efimov resonances $\frac{K_3}{a^4_{\rm{LiCs}}}$ depend on the 3BP, i.e. $\frac{r_{\rm{3b}}}{a_{\rm{CsCs}}}$ since for this system the three-body collisions is closer to being a diabatic process.
$\frac{K_3}{a^4_{\rm{LiCs}}}$ also exhibits St\"uckelberg minima (see the green stripes in the contour plot of Fig.\ref{fig2}(a)) which are virtually insensitive to variations of the ratio $\frac{|a_{\rm{LiCs}}|}{a_{\rm{CsCs}}}$.
This feature of coexisting Efimov resonances and St\"uckelberg minima is inherent in HHL systems with $a_{HL}<0$ and $a_{HH}>0$.
In contrast, for homonuclear recombination processes that produce a shallow dimer, at positive scattering lengths one observes only St\"uckelberg minima.

In the $~^6 {\rm{Li}}-^{133}{\rm{Cs}}-^{133}{\rm{Cs}}$ system the St\"uckelberg minima persist even at $\frac{|a_{\rm{LiCs}}|}{a_{\rm{CsCS}}}<4$.
The latter corresponds to magnetic fields $891\leq B \leq 950~G$ where the ratio $\frac{|a_{\rm{LiCs}}|}{a_{\rm{CsCS}}}\in(4.27,0.05)$ and $a_{\rm{CsCs}}\in(227~a_0,~991~a_0)$ providing enough parameter tunability in order to experimentally probe the St\"uckelberg minima.
From Ref.\cite{berninger_universality_2011}, the 3BR rates are suppressed for Cs atoms at $B=893~G$.
Additional St\"uckelberg physics in the $\rm{Li-Cs}_2$ channel might enable the production of a gas of two bosonic species that keeps high densities throughout a cooling process as was also discussed in Ref.\cite{repp_observation_2013}.

Fig.\ref{fig2}(b, c) compare the scaled 3BR rates calculated with the Eq.~(\ref{eq:eq2}) and with the R-matrix theory, respectively, showing good agreement.
Fig.\ref{fig2}(b, c) exhibit two different log-periodic scaling behaviors versus the axis of $\frac{r_{\rm{3b}}}{a_{\rm{CsCs}}}$ and $\frac{|a_{\rm{LiCs}}|}{a_{\rm{CsCs}}}$, a distinct feature of the Efimov physics in this system.

\begin{figure}[t]
  \centering
  \includegraphics[scale=0.42]{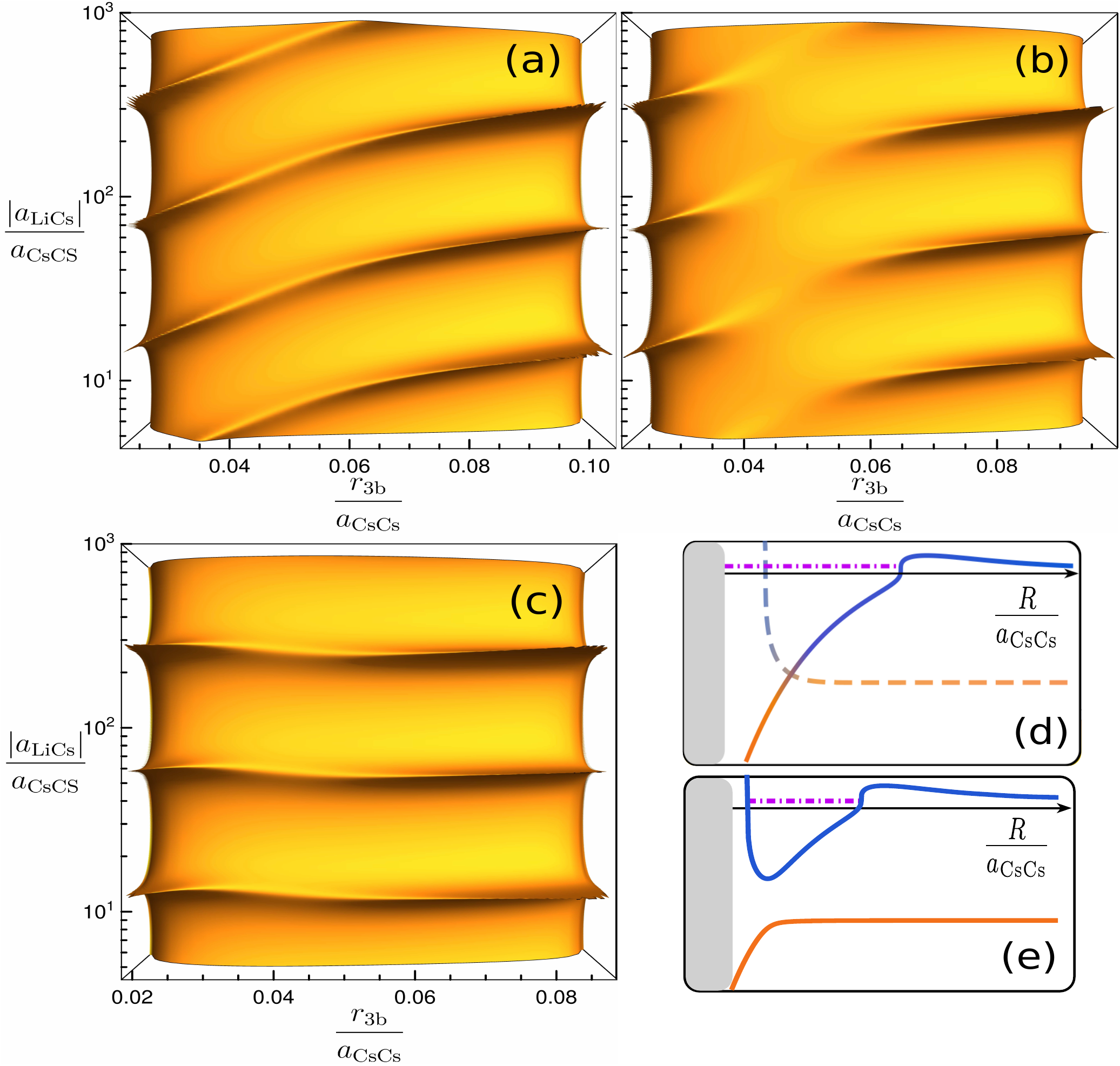}
  \caption{(color online) The scaled 3BR $\frac{m_{\rm{Cs}} K_3}{\hbar a_{\rm{LiCs}}^4}$ at $\bar{E}\to0$ for $~^6 {\rm{Li}}-^{133}{\rm{Cs}}-^{133}{\rm{Cs}}$ using Eq.~(\ref{eq:eq3}).
Panels (a)-(c) correspond to different non-adiabatic probabilities, namely $p=(0.8,~0.4,~0.1)$ respectively.
Panels (d) and (e) present schematic illustrations of the diabatic and adiabatic hyperspherical potentials, respectively. The magenta dash-dotted line indicates the quasi-bound Efimov state and the gray box depicts the 3BP.}
\label{fig3}
\end{figure}

In the spirit of Ref.\cite{dincao_scattering_2005}, the $|S_{12}|^2$ matrix elements in Eq.~(\ref{eq:eq2}) can be further approximated focusing thus on the diabaticity and log-periodicity of the $K_3$. 
When $E$ is the lowest energy scale in the system, the upper and lower adiabatic potential curves illustrated in Fig.\ref{fig1}(a) can be parametrized as $U_1(R)=-\frac{\hbar^2}{2 \mu}\frac{s_0^2+1/4}{R^2}$ (for $a_{HH}\ll R\ll |a_{HL}|$) and $U_2(R)=-\frac{\hbar^2}{2 \mu}\frac{(s_0^*)^2+1/4}{R^2}$ (for $R\ll a_{HH}\ll |a_{HL}|$), respectively.
As in Ref.\cite{wang_universal_2012-2}, $s_0$ ($s_0^*$) corresponds to a universal Efimov scaling coefficient for two (three) resonant interactions.
Then $|S_{12}|^2$ simplifies to: 

\begin{eqnarray}
  &~&  \frac{|S_{12}|^2}{(ka_{HL})^4 }=\frac{p(1-p)\cos^2\bigg(s_0^*\ln \frac{r_{\rm{3b}}}{a_{HH}}+\psi_1\bigg) }{\mathcal{P}(\frac{r_{\rm{3b}}}{a_{HH}},\frac{|a_{HL}|}{a_{HH}})},
  \label{eq:eq3}
\end{eqnarray}

where $\mathcal{P}(\frac{r_{\rm{3b}}}{a_{HH}},\frac{|a_{HL}|}{a_{HH}})=p\sin^2[s_0 \ln \frac{|a_{HL}|}{a_{HH}}+\psi_2-(s_0^*\ln \frac{r_{\rm{3b}}}{a_{HH}}+\psi_1)]+(1-p)\cos^2(s_0 \ln \frac{|a_{HL}|}{a_{HH}}+\psi_2)-p(1-p)\cos^2(s_0^*\ln \frac{r_{\rm{3b}}}{a_{HH}}+\psi_1)$.
$\psi_1$ and $\psi_2$ are arbitrary constant phases and are treated as fitting parameters.
Note that Eq.~(\ref{eq:eq3}) is valid for $|a_{HL}|\gg a_{HH}\gg r_{3B}$.
Elegantly, Eq.~(\ref{eq:eq3}) shows that $|S_{12}|^2$ depends on two geometric scalings where $s_0^*$ ($s_0$) solely determines the St\"uckelberg minima (Efimov resonances).
Note that in the limit of large mass ratio, $s_0^* \approx s_0$.
Fig.\ref{fig3}(a-c) illustrates the effects of diabaticity in $\frac{K_3}{a_{HL}^4}$ using Eq.~(\ref{eq:eq3}) within two successive St\"uckelberg minima where the non-adiabatic probability $p$ is parametrically adjusted, i.e.\ panels (a-c) correspond to $p=(0.8,~0.4,~0.1)$, respectively.
Note that the universal scaling factors $s_0$ and $s_0^*$ are taken from Ref.\cite{wang_universal_2012-2} for the physical system of $~^6 {\rm{Li}}-^{133}{\rm{Cs}}-^{133}{\rm{Cs}}$ whereas the phases $\psi_1$ and $\psi_2$ are arbitrarily chosen.
Evidently, Fig.\ref{fig3}(a-c) unravels the universal aspects of the Efimov resonances in terms of the degree of diabaticity.
Fig.\ref{fig3}(a) depicts the predominantly diabatic regime ($p=0.8$) where the positions of the Efimov resonances strongly depend on the ratio $\frac{r_{\rm{3b}}}{a_{\rm{CsCs}}}$, namely, on the 3BP.
Notice that the {\it{trajectories}\/} of the resonant enhancements, namely {\it{Efimov manifolds}\/}, shift upwards as $\frac{r_{\rm{3b}}}{a_{\rm{CsCs}}}$ increases.
Intuitively, this effect is best understood in the diabatic picture, see Fig.\ref{fig3}(d), where the ``diabatized'' hyperspherical curves are shown whereas the magenta dash-dotted line indicates the Efimov metastable state.
The diabatic potential curve (solid line) which supports the Efimov states depends on the 3BP (gray box).
Therefore, the energy of an Efimov quasi-bound state remains in resonance only by simultaneously increasing the 3BP and the $\frac{|a_{HL}|}{a_{HH}}$.
Note that the maximum of the potential barrier in Fig.\ref{fig3}(e) increases as $R_b\sim 0.34 |a_{HL}| m_H/\mu$. 

Interestingly, in the adiabatic limit, shown in Fig.\ref{fig3}(c) for $p=0.1$, the Efimov resonance manifolds acquire universal characteristics and become virtually independent of the 3BP.
This attribute can be understood by inspecting the potential curves of Fig.\ref{fig3}(e) showing the adiabatic version of the potential curves.
Fig.\ref{fig3}(c) shows that the 3BP only weakly affects the Efimov quasi-bound states (see the magenta dash-dotted line) due to the repulsive barrier at small hyperradius (see the blue solid line) which shields the three particles from exploring the non-universal region that depends on the 3BP.

In the diabatic-to-adiabatic regime, see Fig.\ref{fig3}(b) [$p=(0.4)$], the Efimov manifolds undergo a {\it{rearrangement}\/} where at $\frac{r_{\rm{3b}}}{a_{\rm{CsCs}}}\approx 0.05$ each manifold is ``interrupted'' and for $p=0.1$ [see Fig.\ref{fig3}(c)] is ``reattached''  with the next one.
This effect is mainly related to the widths of the Efimov resonances with respect to $\frac{r_{\rm{3b}}}{a_{\rm{CsCs}}}$.
Particularly, in all the panels of Fig. \ref{fig3} the Efimov manifolds are broader at $\frac{r_{\rm{3b}}}{a_{\rm{CsCs}}}\approx 0.05$ whereas close to the St\"uckelberg minima they become narrower.
This dependence of the width of the Efimov resonances on the 3BP can be understood in terms of the two dominant interfering paths for the recombination into a shallow dimer (see dashed and dot-dashed lines in Fig.\ref{fig1}(a)).
Close to the St\"uckelberg minima due to the destructive interference of the two pathways, the Efimov quasi-bound state is weakly coupled to the atom-dimer continuum, yielding a narrow Efimov manifold.
But for $\frac{r_{\rm{3b}}}{a_{\rm{CsCs}}}\approx 0.05$ the two pathways interfere constructively; hence, the trimer metastable state is strongly coupled to the continuum broadening the Efimov manifolds.
Fig.\ref{fig3}(b) demonstrates that for $p=0.4$ the two pathways interfere maximally yielding maximally broad Efimov manifolds that are completely smeared out by the atom-dimer continuum.

{\it{Conclusions-}} The detailed idiosyncrasies of the Efimov physics for mass-imbalanced ultracold systems are investigated, with a focus on recombination processes into shallow HH dimer states.
Our semiclassical analysis is based on the adiabatic hyperspherical representation including the Landau-Zener physics, yielding a closed-form expression for the corresponding 3BR rate.
A rich Efimov-St\"uckelberg landscape is illustrated from our analysis as a unique feature of the mass-imbalanced systems where the degree of diabaticity constrains the Efimovian universality.
Namely, a diabatic recombination processes depends strongly on the 3BP.
In the diabatic-to-adiabatic regime, the Efimov state manifolds exhibit a rearrangement effect and illustrate the transition from system-dependent Efimov resonances to system-independent ones.
The present development can be generalized to include vdW two-body interactions in order to elucidate the extent to which the 3BP is constrained beyond the zero-range theories.

{\it{Acknowledgments}}- The authors thank P.S. Julienne, S.T. Rittenhouse, N. Mehta and M. Eiles for fruitful discussions and acknowledge the Max Planck Institute for the Physics of Complex Systems in Dresden for the warm hospitality and partial financial support. 
This work was supported in part by NSF grant PHY-1607180. The numerical calculations were performed under NSF XSEDE Resource Allocation No. TG-PHY150003.

\bibliography{few_body.bib}
\end{document}